\documentclass[%
 reprint,
superscriptaddress,
 amsmath,amssymb,
 aps,
]{revtex4-2}
\usepackage{bm}
\usepackage{graphicx}
\usepackage{amssymb}
\usepackage{amsmath}
\usepackage{eufrak}
\usepackage{color}
\usepackage[utf8]{inputenc}
\usepackage[unicode=true,colorlinks=true,urlcolor=blue,citecolor=blue]{hyperref}

\usepackage{pifont}
\usepackage{ulem}
\usepackage[dvipsnames]{xcolor}

\usepackage{graphicx}% Include figure files
\usepackage{dcolumn}% Align table columns on decimal point
\usepackage{bm}% bold math
\usepackage{hyperref}% add hypertext capabilities

\usepackage{orcidlink}

\begin{document}

\preprint{APS/123-QED}

\title{Inverse exciton spin orientation due to trion formation \\ in modulation doped quantum wells}% Force line breaks with \\

\author{Lyubov Kotova~\orcidlink{0000-0001-8767-9252}}\email{kotova@mail.ioffe.ru}
\affiliation{Ioffe Institute, 194021 St.~Petersburg, Russia}
%\affiliation{Faculty of Physics, Lomonosov Moscow State University, 119991 Moscow , Russia}

\author{Alexei Platonov~\orcidlink{0000-0002-7327-2475}} 
\affiliation{Ioffe Institute, 194021 St.~Petersburg, Russia}

\author{Vladimir Kochereshko~\orcidlink{0000-0002-5673-8237}} 
\affiliation{Ioffe Institute, 194021 St.~Petersburg, Russia}

\date{\today}% It is always \today, today,

\begin{abstract}
Time-resolved and time-integrated circularly polarized photoluminescence of excitons and trions in external magnetic fields up to 10~T has been studied in undoped and n-type doped quantum well structures based on ZnSe. In an undoped structure, a circular polarization of photoluminescence induced by magnetic fields corresponded to the Boltzmann distribution of excitons on Zeeman sublevels. The inverse spin orientation of excitons is observed in doped samples with a carrier density of $3 \times 10^{10}$ cm$^{-2}$ and higher. Model calculations show that the reason for the inverse spin orientation is the effective depletion of the lowest-exciton Zeeman sublevel as a result of the spin-dependent formation of trions. The trion formation time as a result of exciton-electron binding was determined as 2~ps. This is noticeably shorter than the characteristic time of the exciton-photon interaction. The observed effect can be considered as a way of spin manipulation by electric fields.
\end{abstract}

\maketitle

%\tableofcontents

%\section{\label{sec:level1}First-level heading:\protect\\ The line break was forced \lowercase{via} \textbackslash\textbackslash}

Spintronics has still confidently attracted great research interest as a possible method of carrier spin manipulation, similar to the carrier current manipulation widely used in semiconductor devices. 

Initially, the idea of controlling the spin current arose in the works on the injection of carrier spin from a magnetic semiconductor into a nonmagnetic semiconductor in the presence of a magnetic field~\cite{1,2,3}. At present, promising directions for the use of spin current in electronics are the use of spin-galvanic effects~\cite{4, 5}. Another direction of spintronics is related to spin storage in quantum dots~\cite{6}. Currently, electron spin storage and spin control appear to be one of the most promising directions for quantum computing~\cite{7}. Combining conventional electronics with spin electronics, is very attractive for the practical utilization of spin phenomena. This can be realized in structures that contain an electron gas. In such structures, it is possible to combine optical and transport phenomena due to the interaction of optically created excitons and free electrons, which leads to trion formation. 
 
In these works, the detection of spin-polarized electrons was based on the observation of circularly polarized exciton and trion emission in non-magnetic single quantum well (SQW) structures. 
 
Charged exciton complexes (trions) containing two electrons and one hole (negatively charged trion) or two holes and one electron (positively charged trion) ~\cite{8} can give us another promising tool for spintronics. Trions are very sensitive to electron spin polarization due to the specific singlet spin structure of the ground state, in which the two electrons have antiparallel spin orientation. It has been shown that trions can be effectively used for detection of electron spins~\cite{9}. Recently, it was proposed to utilize the effects of electron spin flipping processes at trion sublevels in n-type quantum dot structures for optically controlled spin manipulation of electrons~\cite{10}. 
 
In the present work, a detailed study of the inverse spin orientation effect of excitons has been carried out with the aim of using it in spintronics to manipulate spins using electric fields. 
 
ZnSe/Zn$_{0.89}$Mg$_{0.11}$S$_{0.18}$Se$_{0.82}$ QW heterostructures with a SQW thickness of $80\AA$ were grown by molecular beam epitaxy on (100)-oriented GaAs substrates. Most of the structures used in this work were weakly doped. Free electrons in the QW were provided by a 3.0 nm thick doped Cl layer separated from the QW by a 10 nm thick spacer. The electron density could be adjusted from $5 \times 10^9$ to $10^{11}$ cm$^{-2}$ with additional UV illumination. Time-integrated and time-resolved circularly polarized photoluminescence (PL) spectra of excitons and trions were investigated in magnetic fields B up to 10 T in Faraday geometry. 
 
	PL excitation was performed by a synchronously pumped dye laser with energy $E_{ex}=2.8459$ eV, pulse duration $\tau_p=2 ps$ and repetition rate $f=76$ MHz. The average excitation density was $10$ mW/cm$^{-2}$. The PL signal was recorded using a double monochromator in subtractive dispersion mode and a streak camera, giving an overall spectral and temporal resolution of 0.5 meV and 10 ps, respectively. A streak camera was used for time-resolved measurements and a photomultiplier tube was used for time-integrated measurements. The time constants of exciton and trion decay were obtained from the streak traces by deconvolution.
 
	The electron concentration in the QW was determined using the methods described in~\cite{11, 12}. The electron concentration was estimated for our experiments to be $3 \times 10^{10}$ cm$^{-2}$.

\begin{figure*}[t]
\centering
\includegraphics[width=0.65\linewidth]{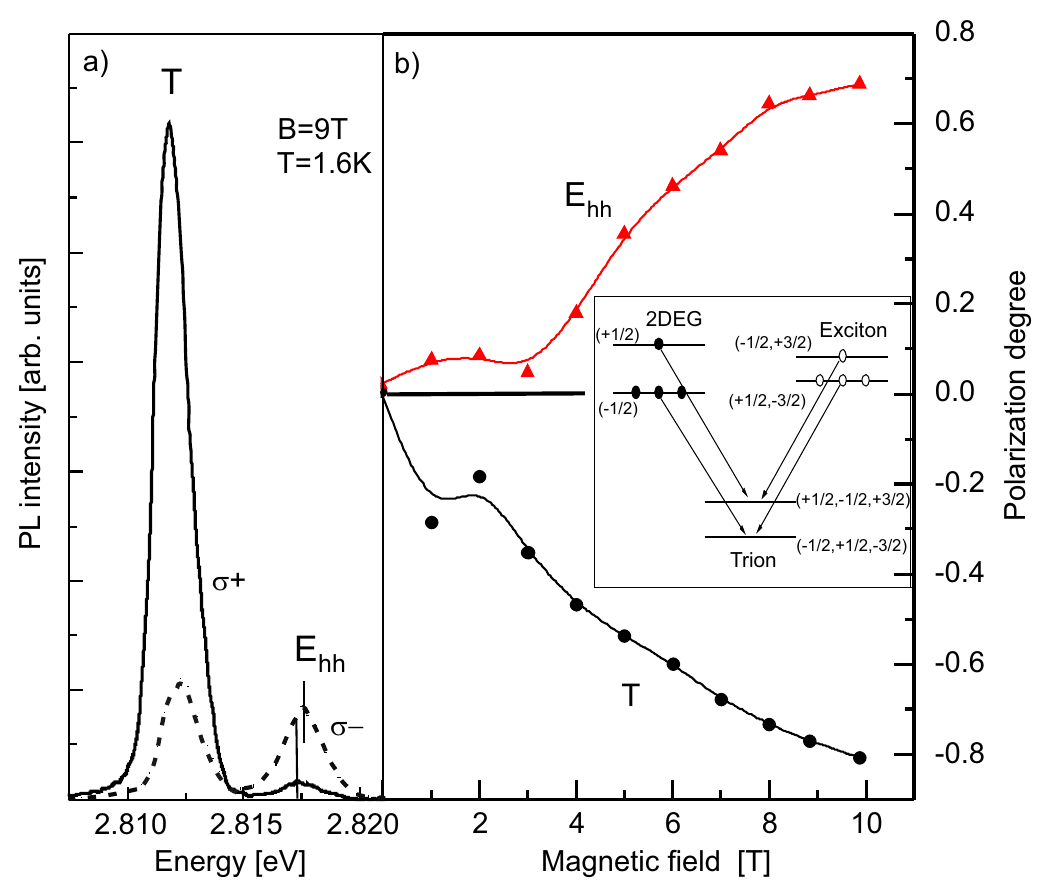}
  \caption{ \textbf{(a)}  Photoluminescence spectra of ZnSe/Zn$_{0.89}$Mg$_{0.11}$S$_{0.18}$Se$_{0.82}$ SQW structure with electron concentration $3 \times 10^{10}$ cm$^{-2}$ taken in magnetic field of $9$ T in two circular polarizations $\sigma^+$  and $\sigma^-$  at $T=1.6$ K. 
\textbf{(b)} Magnetic field dependence of the degree of circular polarization of PL for exciton $(E_{hh})$ and $(T)$ trion. Non-monotonicity on the curves is related to the intersection of dipole active and dipole forbidden exciton levels (Figure~\ref{fig:3}). In the inset, the scheme of the trion formation by exciton and electron coupling is presented.  }
  \label{fig:1}
\end{figure*}

	 Figure~\ref{fig:1}~a shows time-integrated PL spectra of the QW structure in magnetic field of 9 T in $\sigma^+$  and $\sigma^-$  circular polarizations. Figure~\ref{fig:1}~b presents the degree of circularly polarized PL $P_{\rm c}=\frac{I^{+}-I^{-}}{I^{+}+I^{-}}$ (here $I^+, I^-$  are PL intensity in $\sigma^+$  and $\sigma^-$  circular polarizations) of exciton and trion taken at 1.6 K as a function of applied magnetic fields. The degree of circular polarization of the trion PL increases from 0 to 80\% in magnetic field of 10 T according to the trion thermal distribution on Zeemann sublevels. Magnetic field dependence of the degree of circular polarization of the exciton line is similar to the trion one. But the signs of the exciton and trion polarizations are opposite. 
 
	Comparing the exciton and trion PL in magnetic fields (Figure~\ref{fig:1}~a) one can see that the high-energy Zeemann component of the exciton line is more intense than the low-energy component. At the same time, for the trion Zeemann components we have a normal intensity ratio - the low-energy component is more intense than the high-energy one according with the Boltzmann distribution. This indicates the inverse population of the exciton Zeemann sublevels in magnetic fields. 
 
	The observed inverse population can be explained by the depletion of the lowest exciton Zeemann sublevel due to spin dependent trion formation in modulation doped QW structures~\cite{11}. The scheme of such spin dependent trion formation is shown in the inset to Figure~\ref{fig:1}~b. In the presence of magnetic field the electrons in 2DEG are spin polarized preferably in the spin state $( -1/2)$. Since the trion ground state is a singlet, the trions preferably form from excitons on the lowest Zeemann sublevel with momentum $( 1/2,-3/2)$. If the trion formation rate is high enough, this leads to the depletion of the lower energy exciton Zeemann sublevel and consequently to the inverse population of the exciton Zeemann sublevels. 

 \begin{figure}[h]
\centering
\includegraphics[width=0.9\linewidth]{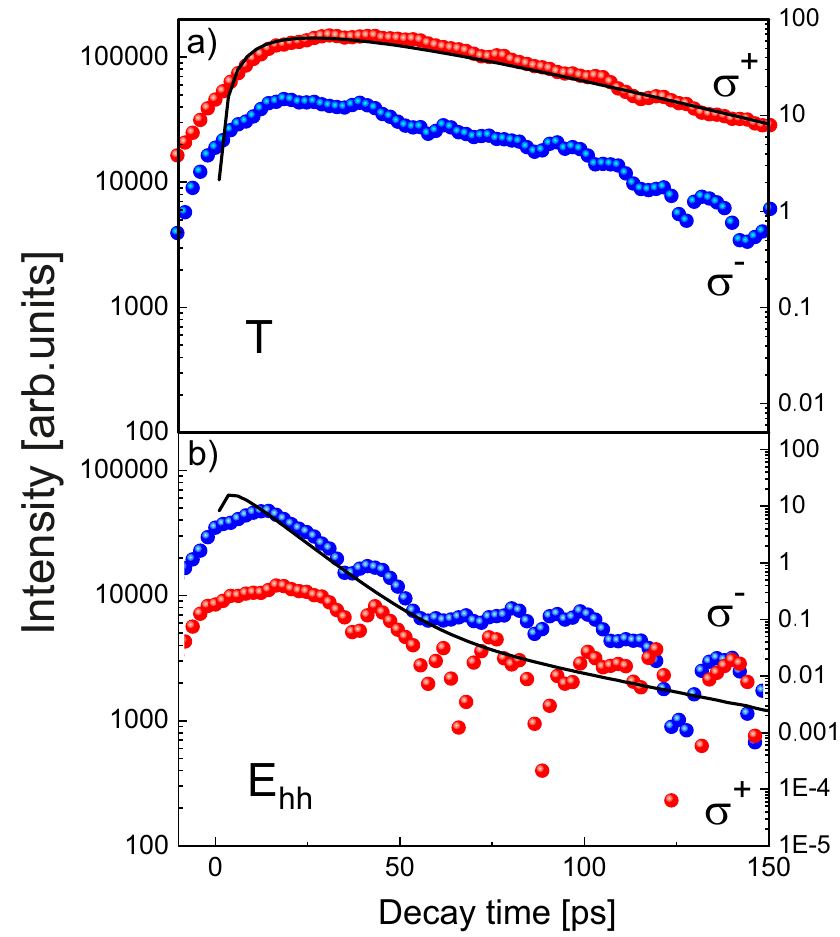}
   \caption{ Decay curves for trion - \textbf{(a)} and exciton - \textbf{(b)} PL line in $\sigma^+$ (red) and $\sigma^-$ (blue) circular polarizations. Solid line - calculation, dotted curve - experiment.}
  \label{fig:2}
\end{figure}

 \begin{figure*}[h]
\centering
\includegraphics[width=0.9\linewidth]{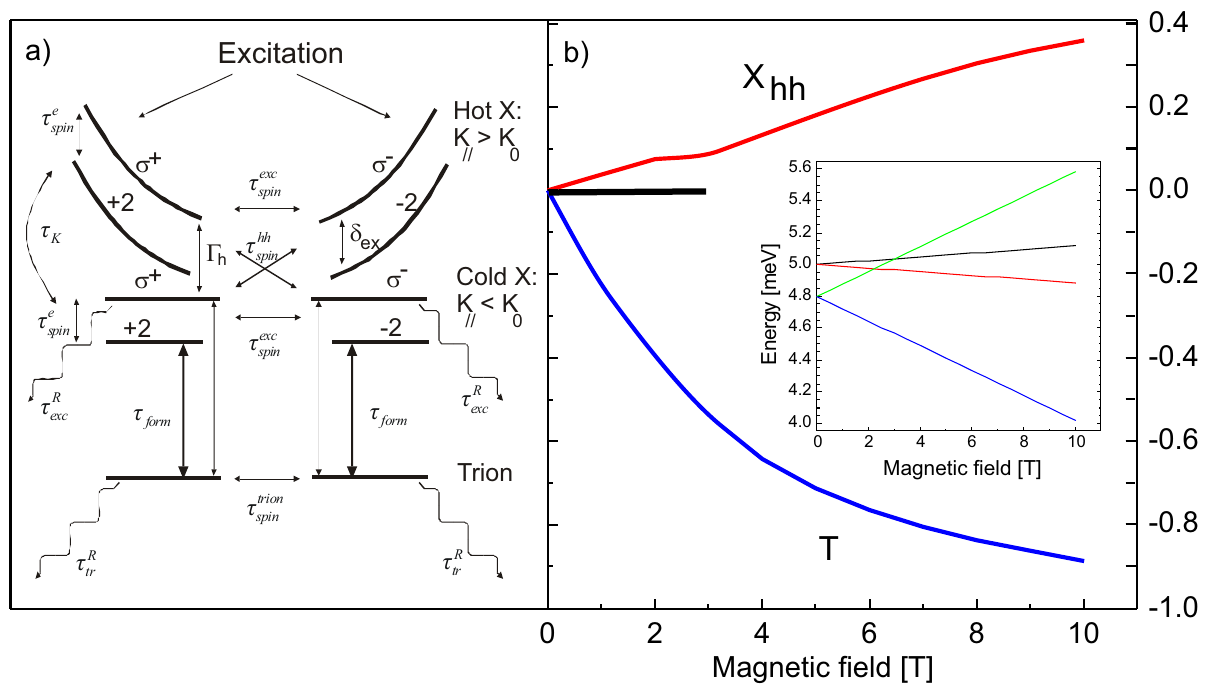}
   \caption{ \textbf{(a)} Schematic representation of the rate equation (see in the text). \textbf{(b)} Calculated magnetic field dependence of the degree of circular polarization for exciton $(E_{hh})$ and $(T)$ trion PL. Calculation parameters: $\tau_K = 15~ps$, $\tau_{tr}^R = 20~ps$, $\tau_{exs}^R=15~ps$, $\tau_{spin}^{exs}=\tau_{spin}^{hh} = 17~ps$, $\tau_{spin}^{trion}=\tau_{spin}^{e} = 40~ps$, $\tau_{form}=2~ps$ measured from the trion level. The insert shows magnetic field dependence of the exciton Zeeman sublevels. }
  \label{fig:3}
\end{figure*} 

We also performed time-resolved measurements of the polarized PL at the temperature of 1.6 K. The exciton and trion PL decay curves taken in magnetic field of 5 T are shown in Figure~\ref{fig:2}~a,b for $\sigma^+$  and $\sigma^-$  circular polarizations. The decay curves are similar for both polarizations. The trion decay is monoexponential with a time constant of 20 ps. One can see that the exciton decay is biexponential. The time constant for the fast exponent is 8 ps and for the slowest one is $\propto20$ ps.

 For the quantitative description of the observed phenomena and understanding of the whole dynamics of the exciton-trion system, model calculations we performed including all possible transitions as it is shown schematically in Figure~\ref{fig:3}~a. At nonresonant optical excitation the excitons with high in-plane $K_\parallel$  vector $(K_\parallel>K_0, K_0$ is the wavevector in vacuum $K_0 = w/c)$ cannot recombine since they are outside the lightcone, but can flip their spin before relaxation to the bottom of the band to states with low   $K_\parallel$ values. The cold excitons $K_\parallel<K_0$  can flip their spin, can be excited to the states with  $K_\parallel>K_0$, can recombine and can form a trion by coupling with an electron having a proper spin. The trions can flip their spin, dissociate into an exciton and electron or can recombine. 
 
	The system of coupled rate equations, which is similar to the system used in~\cite{13, 14} has been solved numerically. 
\begin{equation}
\label{eq1}
	{\partial n_i \over \partial t }= \sum_{J}(n_jw_{ji}-n_iw_{ij})+g_i,
	\nonumber
\end{equation}	
here: $n_i$  is the exciton concentration on the $i$-th level, $w_{ij}$  is the transition rate from the $i$-th level to $j$-th level defined by the relaxation time $w_{ij}=1/2\tau_{ij}$,  $g_i$ is the generation rate for the $i$-th level. The transition rates $w_{ij}$   and  $w_{ji}$  are connected with each other by the relation $ {w_{ij} \over w_{ji}} = e^{\Delta_{ij}\over {kT}};\Delta_{ij}=(E_i-E_j)$, here: $\Delta_{ij}$   is the energy difference between $i$-th and $j$ levels. 

        	The values of g-factors $g_e=1.14$ and $g_{hh}=1.75$ are taken from~\cite{15} (see also~\cite{12}). The exchange splitting between $(\pm 1)$ and $(\pm 2)$ exciton sublevels is taken as $\Delta =0.2$ meV~\cite{16}. 
         
	The system contains eight parameters: exciton and trion lifetimes, exciton and trion spin relaxation times, trion formation time, electron and hole spin relaxation times, and hot exciton relaxation time. Part of them is known and part of them can be obtained from these experimental data. 
 
	We estimate the relaxation time of hot excitons to be $\tau_K =15$ ps from the value of homogeneous exciton linewidth: $ \Gamma_h=1/2\tau_K, \Gamma_h=0.4$~meV~\cite{17}.

	We found that at resonant excitation of the exciton the trion formation time is less than the resolution time of our streak camera $(\tau_{form}<10 ps)$. Consequently, the trion decay curves (Figure~\ref{fig:2}~a) are determined by the radiative lifetime of the trion $\tau^R_{tr} \approx 20$ ps. Due to the exciton localization its radiative lifetime $\tau^R_{exc}$  could not be shorter than the radiative lifetime of free exciton $\tau^R_{exc}>8$ ps and it could not be essentially longer than the trion lifetime~\cite{18}. We choose it in the range $8 ps<\tau^R_{exc} <20 $ ps. The trion spin relaxation time is equal to the hole spin relaxation time. We can estimate $\tau^{hh}_{spin}$  from the saturation level of the magnetic field dependencies of circular polarized PL (Figure~\ref{fig:1}~b) (in the saturation level the degree of polarization is:$P_c = {\tau^R_{tr} \over{\tau^R_{tr}+\tau^{hh}_{spin}}}$ ) to be  $\tau^{hh}_{spin}\propto15-20$ps. The electron spin relaxation time $\tau^{e}_{spin}$  should be longer than the hole spin relaxation time $ \tau^{e}_{spin} > \tau^{hh}_{spin}$~\cite{19}, and we chose  $\tau^{e}_{spin} \approx 40 $ps. We must emphasize here that the result of the calculations is insensitive to these values in the case if: $\tau^R_{tr},  \tau^R_{exc}, \tau^{hh}_{spin} ,\tau^{e}_{spin}  >> \tau_{form}$. 
 
	Because we observe inverse exciton polarization (Figure~\ref{fig:1}~b) we can conclude that the trion formation time is shorter than all the other exciton and trion times. The value  $\tau_{form}=2$ ps gives a satisfactory fit of the calculated and experimentally measured magnetic field dependencies of circularly polarized PL. 
 
	Solid lines in Figure~\ref{fig:2} show the decay curves calculated for PL polarized with exciton and trion. The observed features on the exciton and trion decay curves are well reproducible in our calculations that indicate the correct values of the parameters in the calculation.

The calculated degree of circular polarization of the PL as a function of magnetic fields is shown in Figure~\ref{fig:3}~b. Comparison of Figure~\ref{fig:1}~b and Figure~\ref{fig:3}~b shows reasonable agreement between the calculated and experimental curves. The cusp in the dependence of the exciton degree of polarization at magnetic field of 2 T can be attributed to the crossing of the optically active and dark exciton states shown in the inset in (Figure~\ref{fig:3}~b). A similar cusp is present in the experimental curves (Figure~\ref{fig:1}~b).

\begin{figure}[h]
\centering
\includegraphics[width=0.8\linewidth]{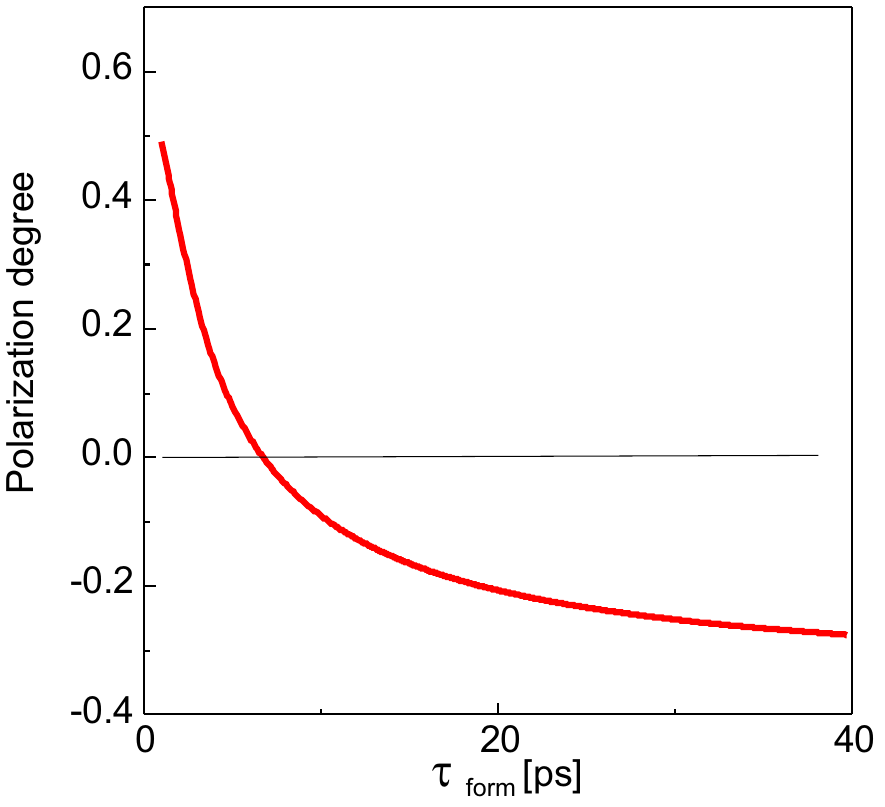}
   \caption{ Degree of circular polarization of the exciton PL as a function of the trion formation time, calculated with the parameters presented in the text.}
  \label{fig:4}
\end{figure}

\begin{figure}[h]
\centering
\includegraphics[width=0.9\linewidth]{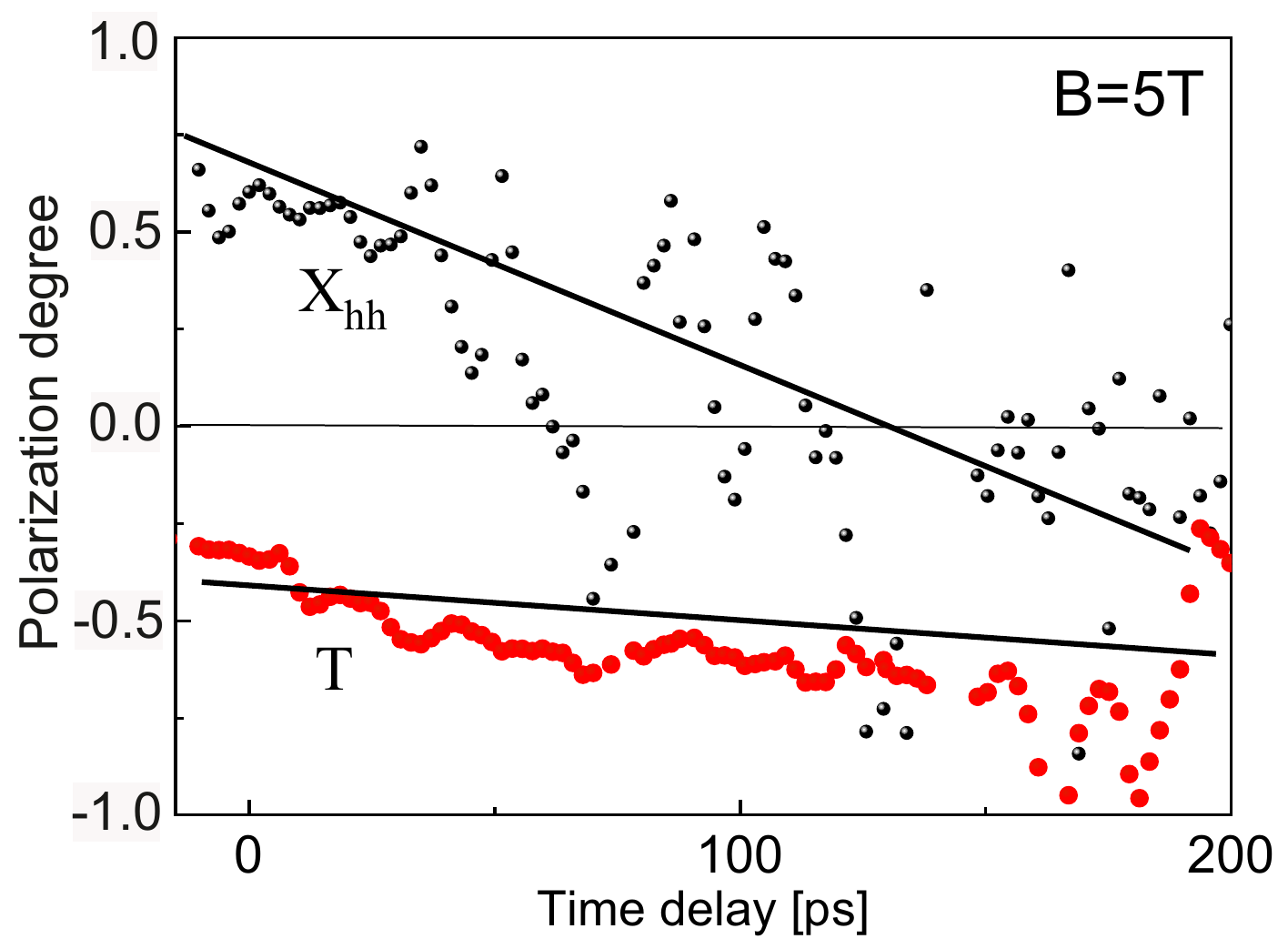}
   \caption{Decay of excitons and trions in a 5~T magnetic field. The dots are experiment. The line is plotted by the least squares method. The large scattering of exciton data is due to the low intensity of this PL line. }
  \label{fig:5}
\end{figure}

The presence of inverse population in the exciton system at their energy relaxation means that this is not yet an equilibrium state and the degree of polarization should change sign once again. Indeed, the change of sign of the spin orientation of excitons is observed at times larger than 100~psec as can be seen from Figure~\ref{fig:5}. 

	We have already mentioned earlier that this negative spin orientation occurs due to the fast spin dependent trion formation. Figure~\ref{fig:4} illustrates the calculated dependence of the degree of circular polarization of the exciton line on the trion formation time for the magnetic field of 10~T. It is clearly seen that the inverse sign of the exciton polarization can be obtained when the trion formation time is shorter than 8~ps. If this time is longer one can see normal sign of the polarization degree. We should mention that in undoped QWs the degree of the exciton PL polarization and consequently the population of the Zeemann sublevels are normal~\cite{12}. 
 
	So, if we would be able to change the trion formation time in a wide range we would be able to control the exciton spin. But the trion formation time depends strongly on the electron concentration and is varied from infinity in undoped QW structures to some small values in heavily doped ones (as it was discussed in~\cite{13}). 
 
	This concentration dependence of the trion formation time can be used for spin manipulation of excitons. Varying the electron concentration by additional UV illumination or by a gate contact one can change the exciton spin polarization from negative to positive values. 
 
	In conclusion: inverse exciton spin orientation in magnetic fields has been observed in modulation doped ZnSe/ZnMgSSe QW structures. Model calculations show that this effect appears due to the depletion of the lowest exciton Zeemann sublevel attributed to the spin dependent trion formation. The value of the trion formation time has been found to be $\tau_{form}=2$ ps. This time is noticeably shorter than the radiation lifetimes of the trion and exciton. The short time of trion formation is also confirmed by the fact that the trion line is clearly manifested in the reflectivity spectra along with the exciton reflection line. We suggest to use the observed effect as a method for exciton spin manipulation by means of electric fields or by optical illumination that is much faster than by magnetic fields.

\begin{acknowledgments}
The authors thank J. Puls for the help with time-resolve experiments. 
\end{acknowledgments}

%\appendix
%\section{Appendixes}

%\begin{verbatim}
%\appendix
%\section{}
%\end{verbatim}
\nocite{*}

\bibliography{apssamp}% Produces the bibliography via BibTeX.

\end{document}